\documentclass[pdflatex]{sn-jnl}

\usepackage{mathrsfs}
\usepackage{graphicx}%
\usepackage{multirow}%
\usepackage{amsmath,amssymb,amsfonts}%
\usepackage{amsthm}%
\usepackage{mathrsfs}%
\usepackage[title]{appendix}%
\usepackage{xcolor}%
\usepackage{textcomp}%
\usepackage{manyfoot}%
\usepackage{booktabs}%
\usepackage{algorithm}%
\usepackage{algorithmicx}%
\usepackage{algpseudocode}%
\usepackage{listings}%
\usepackage{subfig}
\usepackage{booktabs}
\usepackage{float}
\usepackage{chngcntr}

\usepackage[
  backend=biber,
  style=authoryear,
  giveninits=true,
  maxbibnames=30,
  minbibnames=3,
  maxcitenames=1,
  mincitenames=1
]{biblatex}

\DeclareNameAlias{sortname}{family-given}

\DeclareDelimFormat[bib]{multinamedelim}{\addcomma\space}
\DeclareDelimFormat[bib]{finalnamedelim}{\addspace\&\space}
\DeclareDelimFormat[bib]{andothersdelim}{\addspace}

\usepackage{biblatex-shortfields}

\DeclareFieldFormat[article]{title}{#1}

\renewbibmacro*{in:}{%
  \ifentrytype{article}
    {}
    {\printtext{\bibstring{in}\intitlepunct}}%
}

\theoremstyle{thmstyleone}%

\theoremstyle{thmstyletwo}%

\theoremstyle{thmstylethree}%
\newtheorem{definition}{Definition}%

\begin{document}

\title[A Bayesian Spatially Varying Coefficient Model for Surface Urban Heat Island Estimation]{A Bayesian Spatially Varying Coefficient Model for Surface Urban Heat Island Estimation}

%%=============================================================%%

\author*[1]{\fnm{Marika} \sur{D'Agostini}\orcid{https://orcid.org/0000-0001-9998-8896}}\email{marika.dagostini2@unibo.it}

\author[1]{\fnm{Massimo} \sur{Ventrucci}\orcid{https://orcid.org/0000-0001-6903-2076}}\email{massimo.ventrucci@unibo.it}

\author[2]{\fnm{Francesca K.} \sur{de'Donato}\orcid{https://orcid.org/0000-0002-2225-9457}}\email{f.dedonato@deplazio.it}

\author[3]{\fnm{Andrea} \sur{Ranzi}\orcid{https://orcid.org/0000-0002-0037-9167}}\email{aranzi@arpae.it}

\affil[1]{\orgdiv{Department of Statistical Sciences ``Paolo Fortunati"}, \newline \orgname{University of Bologna}, \orgaddress{ 
\city{Bologna}, 
\country{Italy}}}

\affil[2]{\orgdiv{Department of Epidemiology}, \orgname{Lazio Region Health Service}, \orgaddress{
\city{Rome}, 
\country{Italy}}}

\affil[3]{\orgdiv{Environmental Health Reference Centre}, \orgname{Regional Agency for Prevention, Environment and Energy of Emilia-Romagna}, \newline \orgaddress{ 
\city{Modena},  
\country{Italy}}}

%%=============================================================%%

\maketitle
\begin{abstract}

Surface urban heat islands (SUHIs) are commonly studied using satellite-derived land surface temperature (LST), but many analyses rely on static summaries or pointwise urban–rural contrasts, providing limited inference on temporal change, spatial dependence, and uncertainty. This study models LST as a stochastic spatio-temporal process using a spatially varying coefficient formulation that decomposes LST into a baseline spatial field and a location-specific temporal trend. This structure allows persistent thermal differences and local warming rates to be estimated within a unified probabilistic framework. SUHI is derived from the posterior distribution of the baseline spatial field by contrasting the estimated LST at a given pixel with the minimum estimated LST among non-urban pixels. Posterior excursion probabilities are used to identify areas where temporal changes are supported under simultaneous probability control.

The framework is applied to summer Landsat 8 LST observations for Bologna, Italy, from 2013 to 2025. Results show strong spatial heterogeneity in SUHI, with higher values over impervious surfaces and lower values over forests and water bodies. Summer LST increased overall, but local trends varied markedly across the municipality. Joint excursion sets identify areas with warming greater than the domain-wide average and smaller clusters with statistically supported cooling. These results reveal changing thermal patterns across Bologna and help distinguish areas experiencing increasing surface heat from locations where thermal conditions have improved over time. The framework therefore provides an uncertainty-aware basis for monitoring urban thermal change.

\end{abstract}

%%=============================================================%%

\maketitle

%\keywords{Surface Urban Heat Island; Land Surface Temperature; Bayesian Hierarchical Modelling; Spatially-Varying Coefficient Model; Excursion Sets}


%%=============================================================%%

\section{Introduction}\label{sec:Intro}

The urban heat island (UHI) phenomenon refers to areas within a city that record higher temperatures than their non-urban surroundings as a result of the built environment and the prevalence of impervious surfaces \parencite{Oke1982TheIsland}. In studies of the surface urban heat island (SUHI), this contrast is commonly examined using land surface temperature (LST) retrieved from thermal infrared remote sensing \parencite{Stewart2021TheGuidebook}. 

Many SUHI studies characterise thermal contrasts using single-date imagery, selected seasonal scenes, or multi-year averages \parencite{Cheval2024A19912022, Zhou2019SatellitePerspectives}. Although these approaches are useful for describing static spatial patterns, but they do not separate persistent thermal differences from spatially varying temporal change. In addition, spatial and temporal components are often analysed separately, limiting the ability to estimate local warming and cooling rates jointly with baseline thermal conditions \parencite{Diem2024RemotePerspectives, Sinha2025ADynamics}.

The treatment of uncertainty presents a further methodological challenge. Conventional methods can locate areas with relatively high LST, but they generally provide limited information on the joint spatial uncertainty associated with these patterns. Pointwise exceedance probabilities, for example, quantify uncertainty at individual locations but may provide an incomplete description of spatial risk, because neighbouring locations are spatially dependent and pointwise summaries do not account for the joint behaviour of exceedances across contiguous areas \parencite{Cressie2011StatisticsStatistics}.

This study addresses these issues by developing a spatio-temporal Bayesian hierarchical model for SUHI estimation. The model is formulated as a spatially varying coefficient model, with latent spatial processes represented through the stochastic partial differential equation \parencite[SPDE;][]{Lindgren2011AnApproach} approach and integrated nested Laplace approximations \parencite[INLA;][]{Rue2009ApproximateApproximations}. Joint posterior excursion sets are then used to identify areas where temporal warming or cooling is statistically supported 
\parencite{Bolin2015ExcursionModels}.

The proposed approach makes three main methodological contributions. First, the SPDE--INLA formulation represents LST as a continuous spatial process with a Matérn covariance structure while retaining the sparse Gaussian Markov random field representation required for computationally tractable inference on large satellite-derived datasets. This avoids treating individual pixels as spatially independent or requiring aggregation to predefined areal units. Second, the spatially varying coefficient formulation separates persistent spatial variation in baseline LST from spatial variation in the temporal trend, allowing local warming and cooling rates to be estimated within a single probabilistic model. Third, joint posterior excursion sets extend the analysis beyond pointwise posterior summaries by identifying regions where temporal trends credibly exceed specified thresholds while controlling for simultaneous uncertainty across space, thereby linking the fitted spatio-temporal model to probability-based assessments of urban thermal change.

The framework is applied to the municipality of Bologna in northern Italy. Bologna lies at the southern margin of the Po Valley, where the densely built urban area transitions toward agricultural land to the north and the foothills of the Apennines to the south. Its compact historic core is surrounded by residential, commercial, and industrial development, creating marked spatial contrasts in built-up density, vegetation cover, and land-surface characteristics that provide a suitable context for investigating the SUHI phenomenon and the influence of contrasting land-cover patterns on LST. Summer Landsat 8 LST observations from 2013 to 2025 are used to characterise urban surface heat dynamics at high spatial resolution. The analysis examines spatial variation in baseline LST, its relationship with land-cover characteristics, spatial heterogeneity in long-term temperature trends, and areas with statistically supported warming or cooling.

The remainder of the paper is organised as follows. Section 2 describes the data, and Section 3 presents the statistical methodology. Section 4 reports the results for the municipality of Bologna. Section 5 discusses the main findings and limitations, and Section 6 provides conclusions and directions for further work.

\begin{figure}[hb!]
    \centering
    \includegraphics[width=\linewidth]{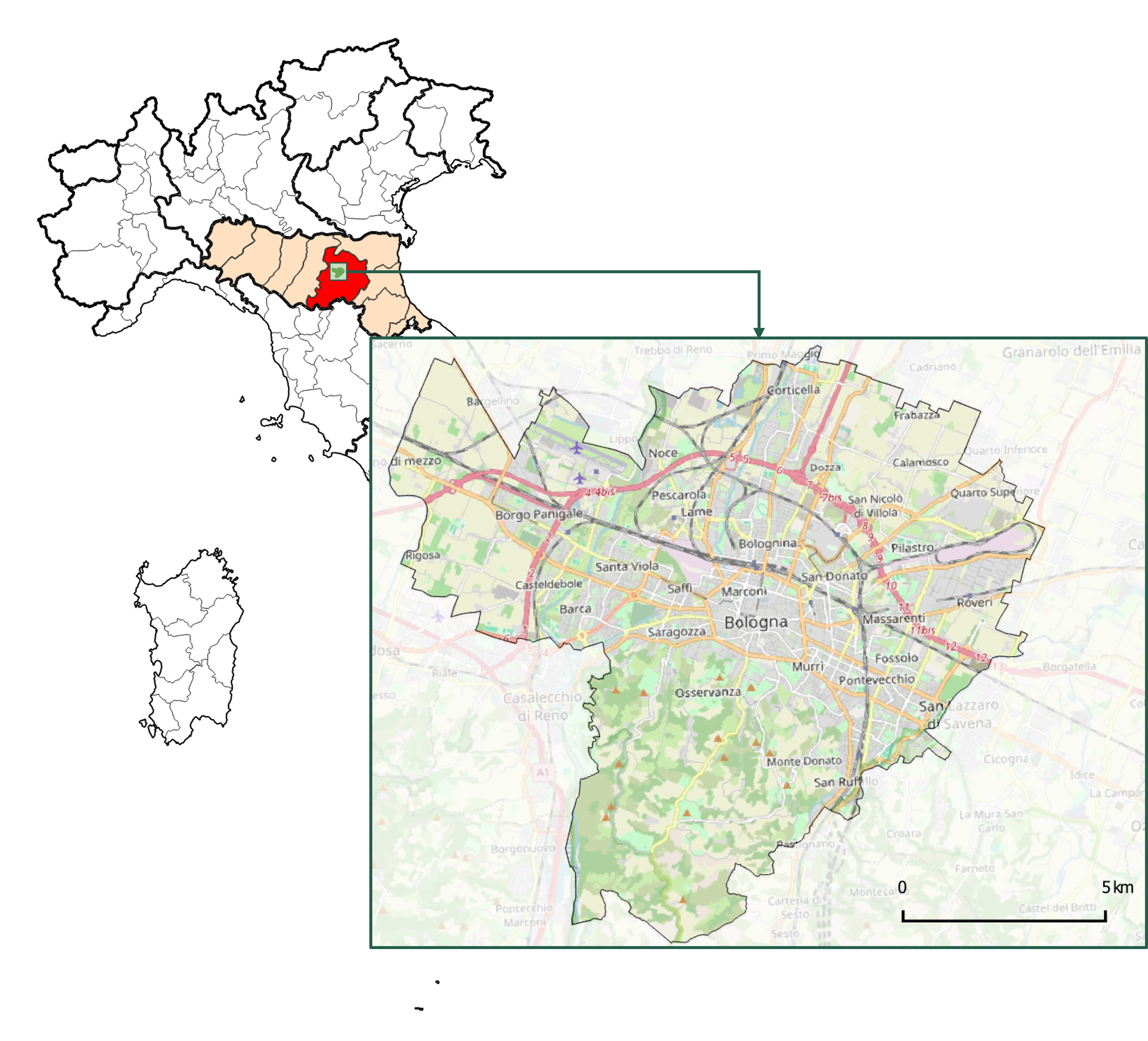}
    \caption{Map of the municipality of Bologna, Italy \parencite{OpenStreetMapData}}
    \label{fig:mapBO}
\end{figure}

\section{Data}\label{sec:Data}

LST data were obtained from the Landsat 8 Collection 2, Level 2, Tier 1 archive through the Google Earth Engine platform \parencite{EarthResourcesObservationandScienceEROSCenter2020LandsatSurvey}. LST was derived from the thermal infrared band ST\_B10, which is provided as an atmospherically corrected surface temperature product by the United States Geological Survey and has a spatial resolution of 30 m. For the municipality of Bologna (Figure~\ref{fig:mapBO}), Landsat 8 observations are available from March 2013 onwards, with a 16-day revisit cycle and a typical overpass time of approximately 10:30 AM local time in winter (CET) and 11:30 AM in summer (CEST).

The analysis uses all available summer scenes, defined as June to August, acquired between 2013 and 2025. Before statistical modelling, each image was processed in three steps. First, scenes were clipped to the 2025 administrative boundary of the municipality of Bologna, obtained from the geographic data warehouse of the Italian National Institute of Statistics \parencite[ISTAT;][]{IstitutoNazionalediStatisticaISTAT2025ConfiniStatistici}. Second, digital numbers from the ST\_B10 band were converted to surface temperature in Kelvin using the linear rescaling coefficients reported in the product metadata:
\begin{equation}
L_{\lambda} = M_L \cdot Q_{DN} + A_L ,
\end{equation}
where $L_{\lambda}$ denotes LST in Kelvin, $Q_{DN}$ is the ST\_B10 digital number, $M_L = 0.00341802$ is the multiplicative scaling factor, and $A_L = 149.0$ is the additive scaling factor. The resulting values were then converted to degrees Celsius. Third, clouds and cloud shadows were removed using the QA\_PIXEL quality assessment band, so that only clear-sky observations were retained. The final LST dataset comprises 69 scenes with at least one clear-sky pixel within the municipal boundary. These scenes yielded 9,072,062 valid spatio-temporal observations over the study period. The temporal distribution of partially clear scenes by year and month, together with summaries of the number of clear-sky pixels and the pixel-level mean, standard deviation, and distribution of observed LST across years and months, is reported in Online Resource 1.

\begin{figure}[hbtp!]
    \centering
    \includegraphics[width=\linewidth]{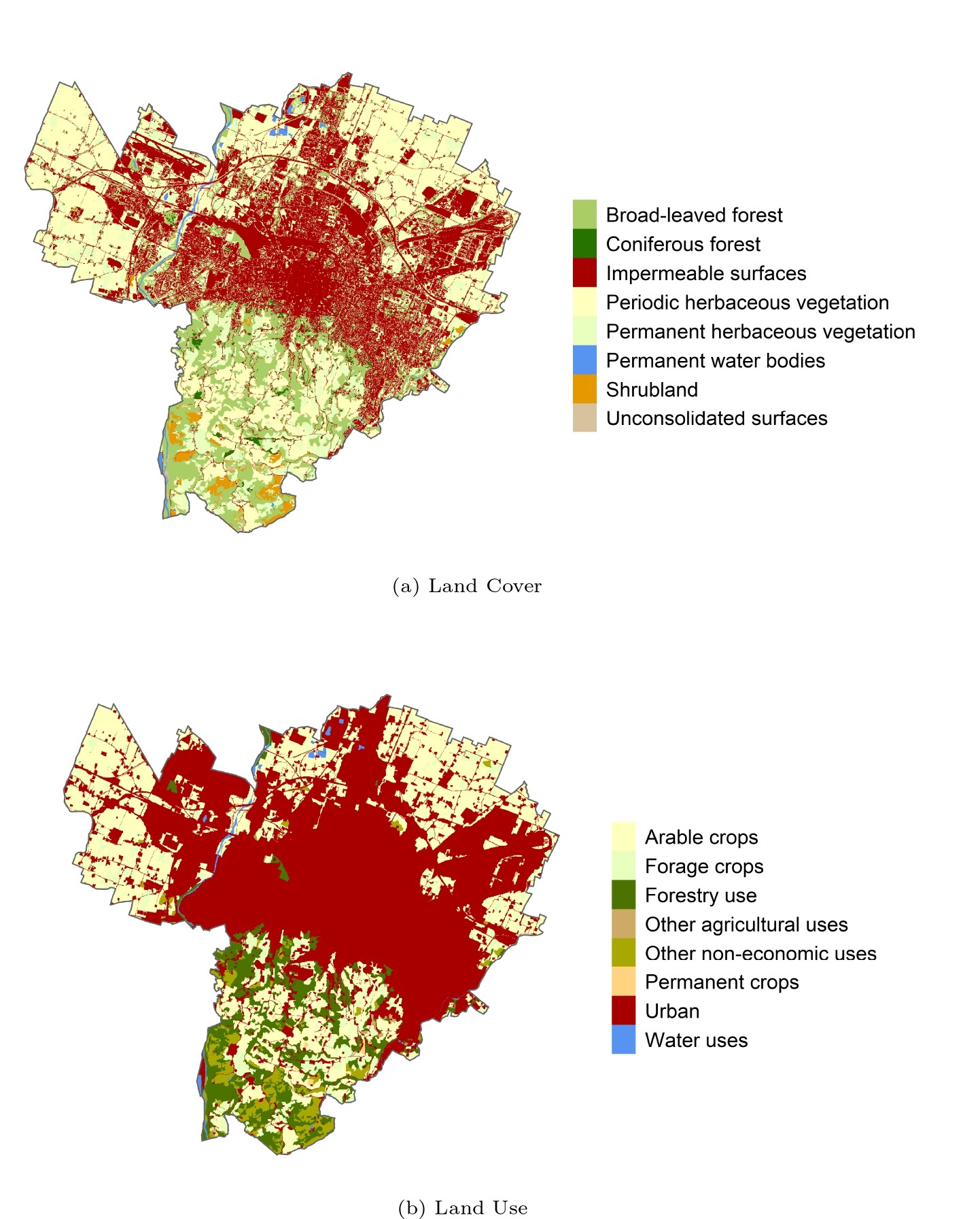}
    \caption{a) Land cover and b) land use maps of the municipality of Bologna (2023)}
    \label{fig:land-use-cover}
\end{figure}

Land cover and land use data were taken from the 2023 national registries produced by the Italian Institute for Environmental Protection and Research \parencite[ISPRA;][]{IstitutoSuperioreperlaProtezioneelaRicercaAmbientaleISPRA2025UsoSuolo}. The datasets follow the CORINE Land Cover classification system and are provided at 10 m spatial resolution. Figure~\ref{fig:land-use-cover} shows the spatial distributions of land cover and land use classes across the municipality of Bologna.

\begin{figure}[htbp!]
    \centering
    \includegraphics[width=\linewidth]{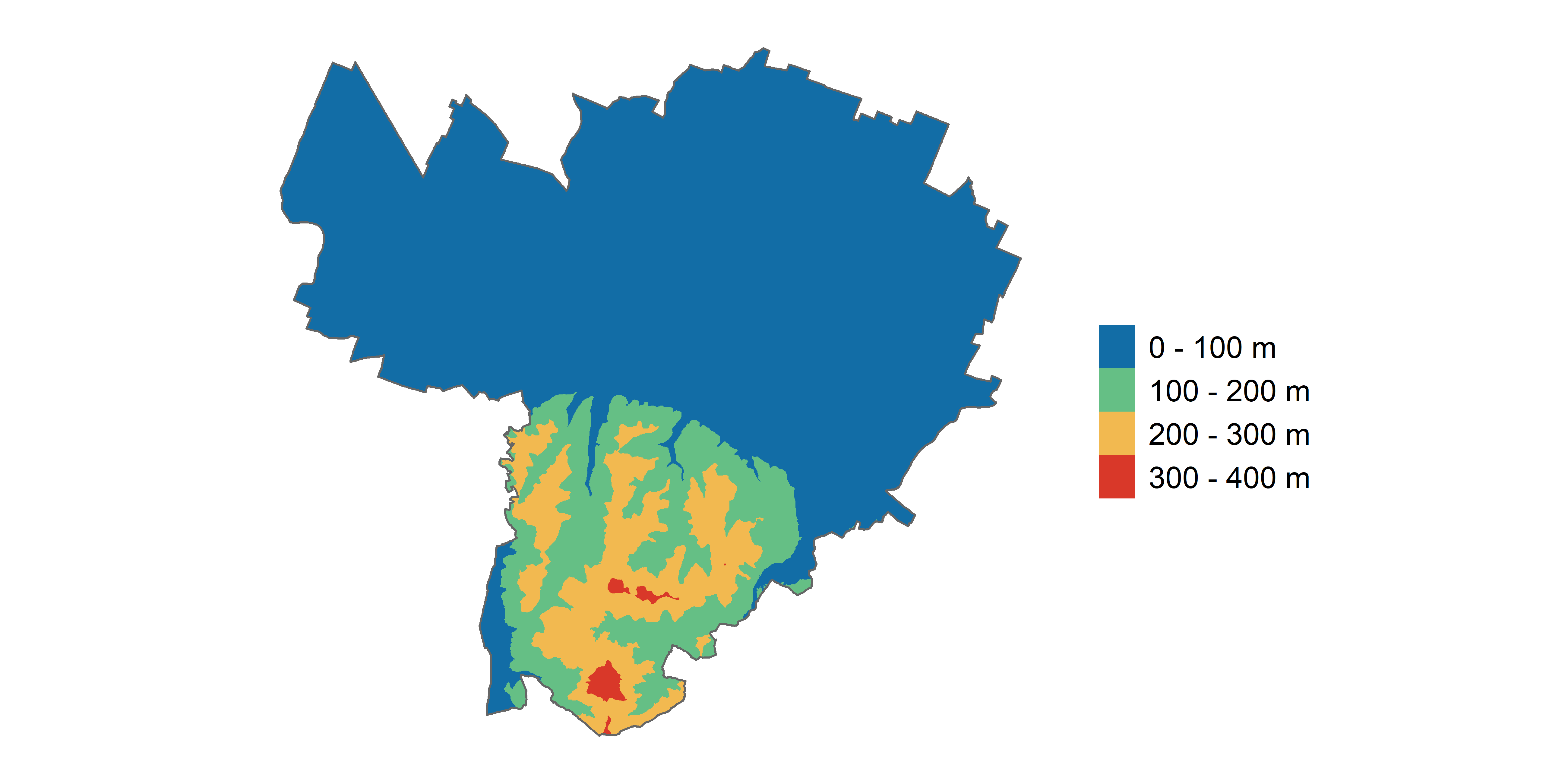}
    \caption{Map of altitude classified into 100 m elevation bands, derived from the resampled Digital Terrain Model (10 m resolution)}
    \label{fig:altitude_classes}
\end{figure}

Topographic elevation was obtained from the Digital Terrain Model (DTM) distributed by the official Geoportal of the Emilia--Romagna Region \parencite{RegioneEmilia-Romagna2025DigitalM}, with a spatial resolution of 5 m. To match the 10 m resolution of the land cover and land use layers, the DTM was resampled by mean aggregation. The continuous elevation surface was then classified into 100 m elevation bands, as shown in Figure~\ref{fig:altitude_classes}, for use in the subsequent definition of the SUHI.

\section{Methodology}

This section presents the statistical framework used to estimate SUHI, spatially varying temperature trends, and areas with statistically supported warming or cooling. We assume a Bayesian spatially varying coefficient model \parencite{Franco-Villoria2019AModels, Gelfand2003SpatialProcesses} with spatial component modelled through the SPDE--INLA approach. 

All data preprocessing, statistical modelling, and spatial analyses were carried out in R, version 4.4.2 \parencite{R2024}. The Bayesian hierarchical model was fitted using the open-source \texttt{R-INLA} project \parencite{Lindgren2015BayesianR-INLA}.

\subsection{Bayesian spatially varying coefficient model}

Let $y(\boldsymbol{s},t)$ denote the observed LST at spatial location $\boldsymbol{s} \in D$ and acquisition time $t \in \mathcal{T}$, where the spatial location corresponds to the centroid of a pixel in the remote-sensing image and time is expressed in centred years. 

The spatial domain $D$ corresponds to the administrative boundary of the municipality of Bologna, with a total of 158,094 pixels, and $\mathcal{T}$ denotes the summer observation period from 2013 to 2025. Conditional on the latent mean $\eta(\boldsymbol{s},t)$ and the observation precision $\tau_{\varepsilon}$, the observations are modelled as:
\begin{align}
y(\boldsymbol{s},t) \mid \ &\eta(\boldsymbol{s},t), \tau_{\varepsilon}
\sim N\left(\eta(\boldsymbol{s},t), \ \tau_{\varepsilon}^{-1}\right)  \\
\eta(\boldsymbol{s},t)
& = \alpha + u(\boldsymbol{s}) + \left[\beta + v(\boldsymbol{s})\right]t . 
\label{eq:svc_model}
\end{align}

Since the time index $t$ is expressed on a continuous yearly scale and centred before model fitting, the global intercept $\alpha$ represents the average baseline LST  over the study domain at the centred reference time, while $\beta$ denotes the global temporal trend, or warming yearly rate. The spatial field $u(\boldsymbol{s})$ describes persistent local deviations from the global baseline $\alpha$, and $v(\boldsymbol{s})$ describes local deviations from the global trend $\beta$. Thus, $\alpha + u(\boldsymbol{s})$ represents the baseline spatial pattern of LST, whereas $\beta + v(\boldsymbol{s})$ gives the location-specific rate of temporal change. This spatially varying coefficient specification separates persistent differences in surface temperature (quantified by the spatial random field $u(\boldsymbol{s})$) from spatial variation in warming rate (quantified by $v(\boldsymbol{s})$), two components that are often combined in descriptive SUHI analyses.

The latent fields are constrained to have zero average over the study domain:
\begin{equation}
\sum_{\boldsymbol{s} \in D} u(\boldsymbol{s}) = 0, \ \ \ \ \ 
\sum_{\boldsymbol{s} \in D} v(\boldsymbol{s}) = 0.
\end{equation}

Under these constraints, $\alpha$ and $\beta$ retain their interpretation as domain-wide average effects, while $u(\boldsymbol{s})$ and $v(\boldsymbol{s})$ represent local departures from these averages.

\subsection{SPDE representation and Penalised Complexity priors}

The spatial fields $u(\boldsymbol{s})$ and $v(\boldsymbol{s})$ are specified as zero-mean Gaussian fields with Matérn covariance functions. For a generic zero-mean spatial field $x(\boldsymbol{s})$, the Matérn covariance between two locations $\boldsymbol{s}_i$ and $\boldsymbol{s}_j$ separated by Euclidean distance $h = || \boldsymbol{s}_i - \boldsymbol{s}_j||$ is defined as
\begin{equation}\label{eq:matern}
\boldsymbol{C}_{\nu}(h; \sigma, \rho)
= \sigma^2 \frac{2^{1-\nu}}{\Gamma(\nu)}
\left(\frac{\sqrt{8\nu}}{\rho} h\right)^{\nu}
K_{\nu}\left(\frac{\sqrt{8\nu}}{\rho} h\right),
\end{equation}
where $\sigma^2$ is the marginal variance, $\nu$ controls smoothness, and $K_{\nu}$ is the modified Bessel function of the second kind \parencite{Stein1999InterpolationKriging}. The practical range is given by $\rho = \frac{\sqrt{8\nu}}{\kappa}$, where $\kappa$ controls the rate of spatial correlation decay, and corresponds to the distance at which the spatial correlation is approximately $0.13$ \parencite{Krainski2019AdvancedINLA}. 

For high-resolution satellite-derived LST data, direct inference with dense covariance matrices is computationally demanding. The Gaussian fields are therefore represented through the SPDE approach \parencite{Lindgren2011AnApproach}. The spatial domain is discretised by a triangular mesh, over which the continuous fields are approximated using finite-element basis functions. This representation yields Gaussian Markov random fields with sparse precision matrices, allowing computationally efficient approximate Bayesian inference with INLA \parencite{Rue2009ApproximateApproximations}. 

\begin{figure}[!b]
    \centering
    \includegraphics[width=\linewidth]{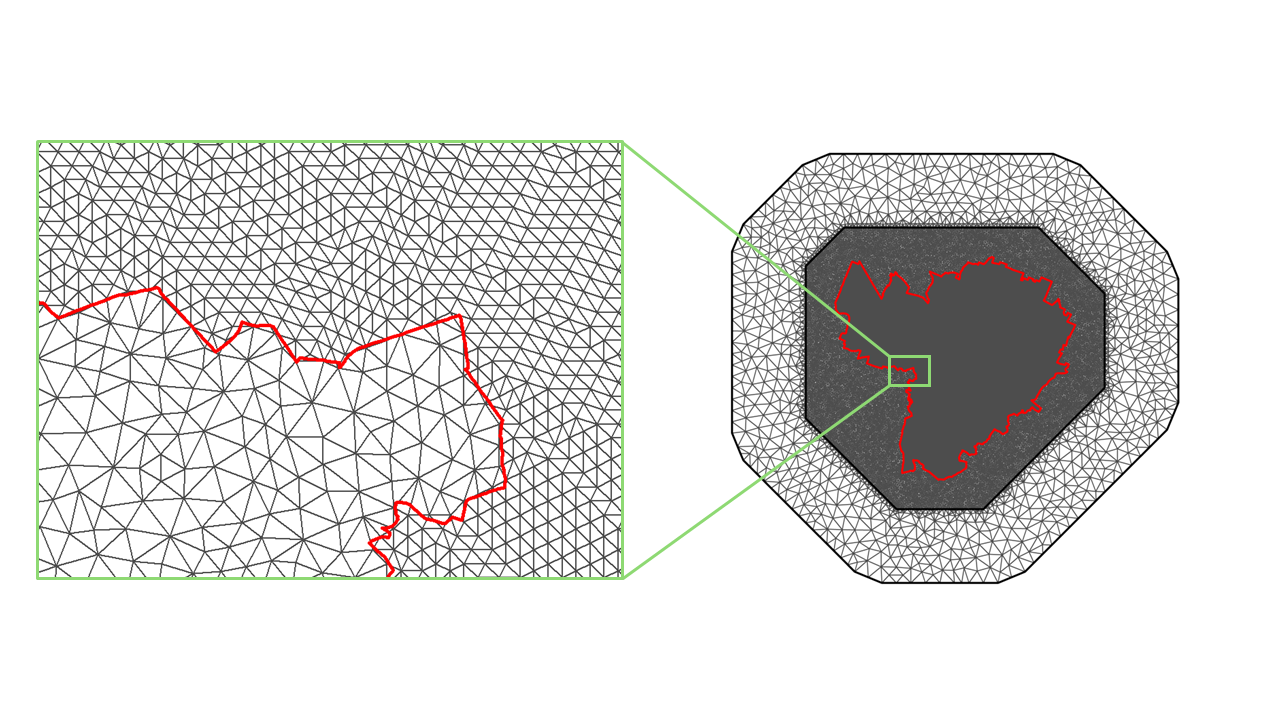}
    \caption{Triangulated mesh used for the SPDE model, with the study area boundary highlighted in red}
    \label{fig:mesh}
\end{figure}

The mesh used in this study has maximum edge lengths of 200~m within the study area and 1200~m in the outer extension, with the former corresponding to approximately 6--7 pixels at the 30~m spatial resolution of the remote-sensing data (Figure~\ref{fig:mesh}). A cut-off distance of 60~m was used to prevent the formation of unnecessarily small triangles around closely spaced input locations, while inner and outer offsets of 2~km and 5~km, respectively, extended the mesh beyond the municipal boundary to reduce boundary effects.

Separate SPDE models are specified for the baseline field $u(\boldsymbol{s})$ and the spatially varying trend field $v(\boldsymbol{s})$. Both fields are assumed to be zero-mean Matérn Gaussian fields, but they are allowed to have different practical ranges and marginal variances:
\begin{align*}
u(\boldsymbol{s}) &\sim \operatorname{GF}\left(0, \boldsymbol{C}_u(\rho_u,\sigma_u^2)\right), \\
v(\boldsymbol{s}) &\sim \operatorname{GF}\left(0, \boldsymbol{C}_v(\rho_v,\sigma_v^2)\right).
\end{align*}
Here, $\boldsymbol{C}_u$ and $\boldsymbol{C}_v$ denote Matérn covariance functions of the form given in Equation \ref{eq:matern}, with $\nu=1$ and hyperparameters $(\rho_u,\sigma_u^2)$ and $(\rho_v,\sigma_v^2)$, respectively. This specification allows the persistent baseline LST field and the spatially varying temporal trend field to operate over different spatial scales and with different magnitudes of spatial variability.

The hyperparameters of the spatial fields are assigned joint penalised complexity (PC) priors \parencite{simpson2017penalising}. These priors shrink the spatial effects towards a simpler base model specification unless the data support additional spatial complexity. Let $d$ denote the dimension of the spatial domain; in this application, $d=2$. Following \textcite{fuglstad2019constructing}, the joint PC prior corresponding to a base model with practical range $\rho = \infty$ and marginal standard deviation $\sigma = 0$ is given by
\begin{align*}
\pi(\rho,\sigma)=\frac{d}{2}
\lambda_{\rho} \lambda_{\sigma}\rho^{-d/2-1}\exp\left(-\lambda_{\rho}\rho^{-d/2}-\lambda_{\sigma}\sigma\right), \qquad\sigma>0, \quad \rho>0
\end{align*}
where the rate parameters $\lambda_{\rho}$ and $\lambda_{\sigma}$ are specified through the probability statements 
\begin{equation}
\mathbb{P}(\rho < \rho_0)=\alpha_{\rho},
\qquad
\mathbb{P}(\sigma > \sigma_0)=\alpha_{\sigma}.
\end{equation}
that are achieved by
\begin{equation}
\lambda_{\rho} = -\log(\alpha_{\rho}) \rho_0^{d/2},
\qquad
\lambda_{\sigma} = -\frac{\log(\alpha_{\sigma})}{\sigma_0}.
\end{equation}

Specifically, for the baseline spatial field $u(\boldsymbol{s})$, the prior is specified as
\begin{equation}\label{eq:pc_u}
\mathbb{P}(\rho_u < 5 \text{ km}) = 0.95, \qquad
\mathbb{P}(\sigma_u > 5 \text{ °C}) = 0.05,
\end{equation}

while for the spatially varying temporal trend $v(\boldsymbol{s})$, the prior is specified as
\begin{equation}\label{eq:pc_v}
\mathbb{P}(\rho_v < 1 \text{ km}) = 0.95, \qquad
\mathbb{P}(\sigma_v > 1 \text{ °C/year}) = 0.05.
\end{equation}
These settings reflect the assumption that baseline LST may vary over broader spatial scales, whereas temporal trend heterogeneity may occur over shorter distances.

The intercept $\alpha$ is assigned an improper flat prior, while the temporal slope $\beta$ is assigned a diffuse Gaussian prior. The observation precision
$\tau_{\varepsilon}$ is assigned a Gamma prior:
\begin{equation}
\pi(\alpha) \propto 1,
\qquad
\beta \sim \mathcal{N}(0,\tau_{\beta}^{-1}), \qquad \tau_{\varepsilon} \sim \text{Gamma}(a_{\varepsilon}, b_{\varepsilon}).
\end{equation}
The hyperparameters are set to $\tau_{\beta}=0.001$, $a_{\varepsilon}=1$, and $b_{\varepsilon}=0.00005$.

A prior sensitivity analysis was conducted for the spatial range parameters of both latent fields. The likelihood, mesh specification, and marginal standard deviation priors were held fixed, while alternative thresholds for the PC priors on the spatial ranges were examined. For the baseline field $u(\boldsymbol{s})$, values $\rho_{0,u}\in\{1,5,10\}$~km were considered under $\mathbb{P}(\rho_u < \rho_{0,u})=0.95$. For the trend field $v(\boldsymbol{s})$, thresholds $\rho_{0,v} \in {0.5,1,5}$ km were considered under $\mathbb{P}(\rho_v < \rho_{0,v})=0.95$. The marginal standard deviation priors were kept fixed at $\mathbb{P}(\sigma_u>5)=0.05$ and $\mathbb{P}(\sigma_v>1)=0.05$. Each specification was fitted independently and compared using the deviance information criterion (DIC), the Watanabe-Akaike information criterion (WAIC), and posterior summaries of the fixed effects and spatial hyperparameters. The analysis was used to assess whether the main substantive conclusions were sensitive to the choice of spatial range priors.

\subsection{Estimation of surface urban heat island}

SUHI estimation is derived from the posterior distribution of the baseline LST field, $B(\boldsymbol{s})=\alpha+u(\boldsymbol{s})$, representing persistent spatial variation in LST at the centred reference time. In Definition \ref{def:suhi}, SUHI is quantified by contrasting the baseline LST at a given pixel with the minimum baseline LST among non-urban pixels.
To account for topographic gradients, comparisons are restricted within 100 m elevation bands \parencite{Richiardi2025ATechnology}. 

\begin{definition}[SUHI]\label{def:suhi}
Let $D_U$ and $D_N$ denote the sets of urban and non-urban pixels, $h(\boldsymbol{s})$ denote the elevation band containing location $\boldsymbol{s}$, and let $D_N^{h(\boldsymbol{s})}$ be the subset of non-urban pixels in the same band. For each urban location, SUHI is defined as
\begin{equation}\label{eq:SUHI}
\operatorname{SUHI}(\boldsymbol{s}) = B(\boldsymbol{s}) - \min_{\boldsymbol{s}' \in D_N^{h(\boldsymbol{s})}} B(\boldsymbol{s}'), \qquad \boldsymbol{s}\in D_U.
\end{equation}
\end{definition}
Equation \ref{eq:SUHI}  compares each urban pixel with the coolest non-urban pixel within the same elevation band. The use of a minimum-based non-urban reference provides a cool-reference comparison with the lowest estimated non-urban conditions in the study area. The resulting SUHI is therefore interpreted as an elevation-adjusted thermal excess relative to a locally attainable cool non-urban reference, rather than as a contrast with average non-urban or rural background conditions.

Minimum-based references can be sensitive to pixel-level noise, retrieval error, or anomalous values when applied directly to raw satellite observations. In this analysis, this issue is reduced because the metric is computed from the posterior distribution of the latent field rather than from individual observations. The reference temperature is therefore obtained from a smoothed continuous thermal surface, with model-based regularisation and uncertainty propagation. This approach is consistent with studies showing the sensitivity of SUHI estimates to reference-area definitions and with spatial analyses aimed at identifying intra-urban thermal hotspots or local extreme conditions \parencite{Martin2014AnIsland, Sobrino2020AData}.

An estimate of the quantity in Equation \ref{eq:SUHI},  $\widehat{\mathbb{E}} \left[ \operatorname{SUHI}(\boldsymbol{s}) \mid \boldsymbol{y} \right]$, is computed from a sample of the posterior distribution of the persistent spatial variation in LST, $B(\boldsymbol{s})$, obtained via INLA.  Let
\begin{equation}
B^{(m)}(\boldsymbol{s})= \alpha^{(m)}+u^{(m)}(\boldsymbol{s}), \qquad m=1,\ldots,1000,
\end{equation}
denote one particular sample from the posterior of $B(\boldsymbol{s})$. For each posterior sample, the elevation-specific reference was computed as
\begin{equation}
R_h^{(m)} = \min_{\boldsymbol{s}'\in D_N^{h(\boldsymbol{s})}} B^{(m)}(\boldsymbol{s}'),
\end{equation}
and the corresponding SUHI surface obtained as
\begin{equation}
\mathrm{SUHI}^{(m)}(\boldsymbol{s}) = B^{(m)}(\boldsymbol{s}) - R_{h(\boldsymbol{s})}^{(m)}, \qquad \boldsymbol{s}\in D_N^{h(\boldsymbol{s})}.
\end{equation}
The posterior mean SUHI intensity is then estimated by averaging the draw-specific values:
\begin{equation}\label{eq:SUHI-est}
\widehat{\mathbb{E}} \left[ \operatorname{SUHI}(\boldsymbol{s}) \mid \boldsymbol{y} \right] = \frac{1}{M} \sum_{m=1}^{M} \operatorname{SUHI}^{(m)}(\boldsymbol{s}).
\end{equation}
Posterior uncertainty is summarised pointwise through the posterior standard
deviation,
\begin{equation}\label{eq:SUHI-sd}
\widehat{\operatorname{SD}} \left[ \operatorname{SUHI}(\boldsymbol{s}) \mid \boldsymbol{y} \right] = \left[ \frac{1}{M-1} \sum_{m=1}^{M} \left\{ \operatorname{SUHI}^{(m)}(\boldsymbol{s}) - \widehat{\mathbb{E}} \left[ \operatorname{SUHI}(\boldsymbol{s}) \mid \boldsymbol{y} \right] \right\}^{2} \right]^{1/2},
\end{equation}
and through the posterior quantiles
\begin{equation}\label{eq:SUHI-quantiles}
\widehat{Q}_{p}(\boldsymbol{s}) = \operatorname{Quantile}_{p} \left\{ \operatorname{SUHI}^{(1)}(\boldsymbol{s}), \ldots, \operatorname{SUHI}^{(M)}(\boldsymbol{s}) \right\}, \qquad p \in \{0.025,0.975\}.
\end{equation}
The interval $\left[\widehat{Q}_{0.025}(\boldsymbol{s}), \widehat{Q}_{0.975}(\boldsymbol{s}) \right]$ therefore provides a pointwise $95\%$ posterior credible interval for the SUHI intensity at location $\boldsymbol{s}$.

The resulting SUHI estimates are finally summarised spatially and by land-cover class. These summaries are used to assess whether higher SUHI intensities are associated with specific surface configurations, including impervious surfaces, herbaceous vegetation, forested areas, and water bodies. 

\subsection{Excursion probabilities for warming and cooling regions}

Posterior excursion probabilities \parencite{Bolin2015ExcursionModels} are used in this study to identify areas with statistically supported temporal change. Unlike pointwise exceedance maps, which evaluate the threshold condition separately at each location, excursion sets provide simultaneous spatial statements by accounting for dependence in the latent field. Specifically, they identify the largest regions in which the latent process exceeds, or falls below, a specified threshold with a prescribed joint posterior probability \parencite{Bolin2015ExcursionModels}. They therefore provide a principled basis for mapping spatially coherent regions of supported warming or cooling.

In this study, excursion sets are computed as a post-processing step from the joint posterior distribution obtained under the SPDE--INLA model. This allows the analysis to move beyond posterior mean trend maps by identifying spatially coherent regions where evidence of temporal change remains after accounting for spatial dependence and posterior uncertainty.

Negative excursion sets are computed for the local temporal trend, $T(\boldsymbol{s})=\beta+v(\boldsymbol{s})$, to identify regions of absolute cooling. These are defined as areas where the local temporal trend is below zero with joint posterior probability of at least 0.95:
\begin{equation}\label{eq:neg-exc}
\tilde{E}^{-}_{0} = \operatorname*{arg max}_{D \subseteq \mathcal{D}} \{ |D| : \mathbb{P}\left(T(\boldsymbol{s}) < 0,\ \forall \boldsymbol{s} \in D \mid \boldsymbol{y} \right) \geq 0.95 \}.
\end{equation}

Positive excursion sets are computed for the localised trend component $v(\boldsymbol{s})$ to identify areas where the warming rate exceeds the municipality-wide average (i.e., $T(\boldsymbol{s})>\beta$):
\begin{equation}\label{eq:pos-exc}
\tilde{E}^{+}_{0} = \operatorname*{arg max}_{D \subseteq \mathcal{D}} \{ |D| : \mathbb{P}\left( v(\boldsymbol{s}) > 0,\ \forall \boldsymbol{s} \in D \mid \boldsymbol{y} \right) \geq 0.95 \}.
\end{equation}

This distinction is relevant because a location may show absolute warming, with $\beta+v(\boldsymbol{s})>0$, while warming more slowly than the municipal average, with $v(\boldsymbol{s})<0$. All excursion regions are computed using the \texttt{excursions} R package \parencite{Bolin2018CalculatingExcursions}.

\section{Results}\label{sec:results}

\subsection{Model parameters and spatially varying coefficients}

Table~\ref{tab:hyperparam} reports posterior summaries for the fixed effects, observation precision, and SPDE hyperparameters of the fitted spatially varying coefficient model defined in Equation \ref{eq:svc_model}. The posterior mean of the intercept $\alpha$ is 39.63~$^\circ$C, representing the average baseline LST across the study area. The global temporal trend $\beta$ has a posterior mean of 0.14~$^\circ$C per year, with a 95\% credible interval from 0.12 to 0.17~$^\circ$C per year. This result indicates an overall increase in summer LST during the 2013--2025 study period. The Gaussian observation precision is estimated close to 0.06, corresponding to a residual standard deviation of approximately 4.08~$^\circ$C. This indicates that some short-term variability remains after accounting for the persistent spatial field and the linear spatially varying trend, as expected for clear-sky satellite LST observations acquired under varying summer meteorological conditions.

\begin{table}[b]
    \caption{Posterior summaries of model hyperparameters and global parameters. Reported quantities include the Gaussian observation precision, the SPDE parameters for the spatial fields $u(\boldsymbol{s})$ and $v(\boldsymbol{s})$, and the global coefficients $\alpha$ and $\beta$}
    \normalsize
    \centering
    \begin{tabular}{lccccc}
         \toprule
        \textbf{Parameter} & \textbf{Mean} & \textbf{SD} & \textbf{Median} & \textbf{2.5\%} &  \textbf{97.5\%} \\
         \midrule
    $\tau_\varepsilon $ & 0.06 & 2.78$\times 10^{-5}$ & 0.06 & 0.06 & 0.06  \\
    \midrule
    $\rho_u$  & 3.09 & 0.30 & 3.08 & 2.53 & 3.69 \\
    $\sigma_u$  & 8.13 & 0.77 & 8.11 & 6.67 & 9.72  \\
    \midrule
    $\rho_v $  & 1.91 & 0.29 & 1.93 & 1.32 & 2.40 \\
    $\sigma_v$  & 0.31 & 0.05 & 0.31 & 0.21 & 0.38  \\
    \midrule
    $\alpha$ & 39.63 & 0.22 & 39.63 & 39.20 & 40.06  \\
    $\beta$ & 0.14  & 0.01 & 0.14 & 0.12  & 0.17 \\
    \bottomrule
           \end{tabular}
    \label{tab:hyperparam}
\end{table}

\begin{figure}[htbp!]
  \centering
  \includegraphics[width=\textwidth]{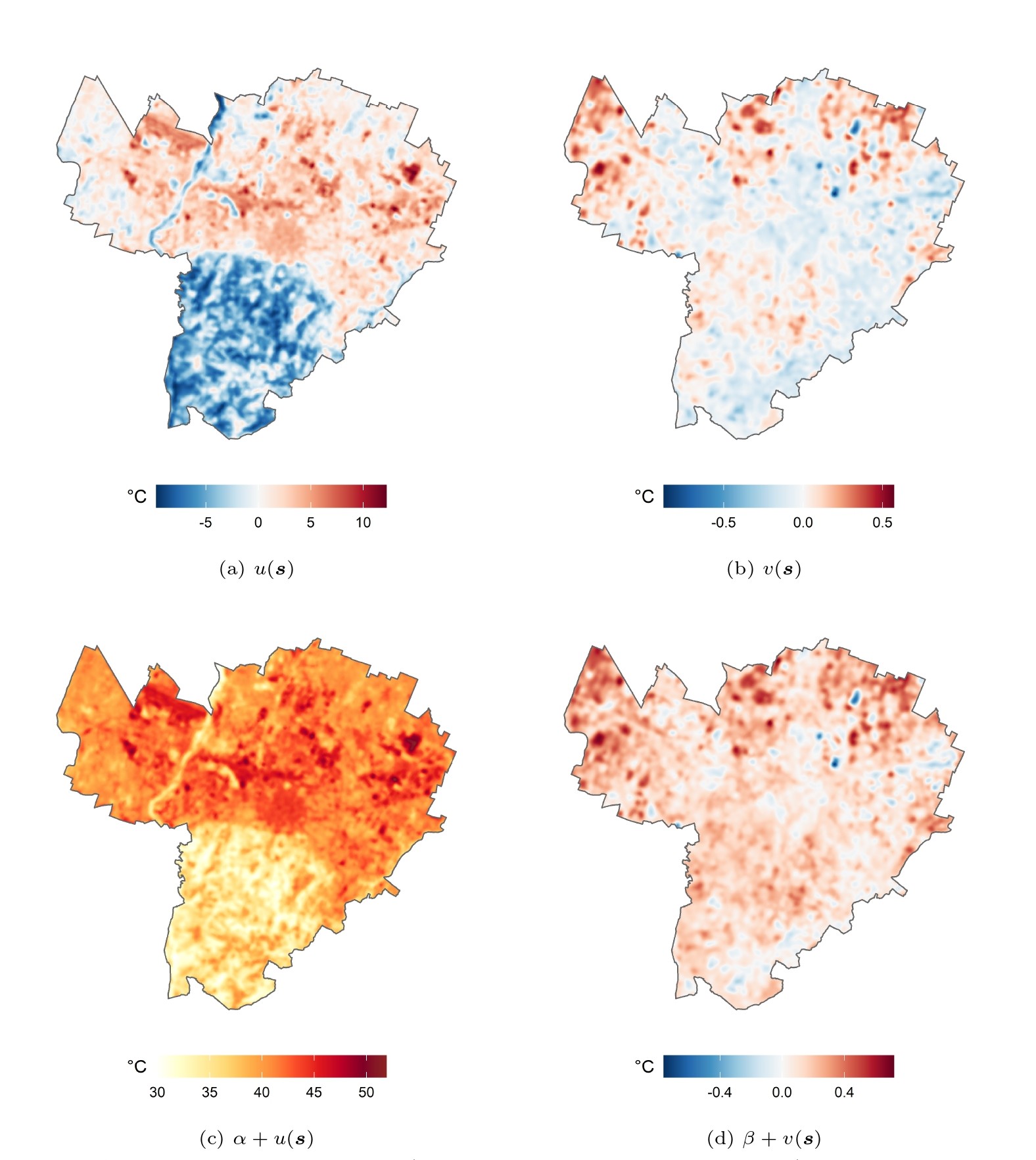}
  \caption{Posterior means of a) the spatial field $u(\boldsymbol{s})$, b) the spatial field $v(\boldsymbol{s})$, c) the persistent spatial effect, and d) the local rate of change in land surface temperature}
  \label{fig:posterior_mean_u_v}
\end{figure}

Figure~\ref{fig:posterior_mean_u_v} shows the posterior mean surfaces for $u(\boldsymbol{s})$, $v(\boldsymbol{s})$, $\alpha+u(\boldsymbol{s})$, and $\beta+v(\boldsymbol{s})$. The posterior standard deviation of $u(\boldsymbol{s})$ ranges from 0.32 to 0.61~$^\circ$C, while that of $v(\boldsymbol{s})$ ranges from 0.03 to 0.07~$^\circ$C per year (see Online Resource 2 for maps). The posterior correlation between $u(\boldsymbol{s})$ and $v(\boldsymbol{s})$ is approximately $-0.08$. This weak correlation indicates that the two latent fields describe largely distinct components of spatial variation. Areas with high baseline LST do not necessarily show stronger temporal warming, and areas with lower baseline temperatures do not necessarily show weaker or negative trends.

\subsection{Baseline land surface temperature by land cover and land use}

To examine the association between surface characteristics and baseline temperature, posterior mean values of $\alpha+u(\boldsymbol{s})$ were summarised across the 2023 land cover and land use categories shown in Figure~\ref{fig:land-use-cover}. The thematic layers are available at 10~m resolution, whereas the Landsat-derived LST observations are defined on a 30~m grid. For this reason, posterior summaries of the continuous spatial field were projected from the SPDE mesh onto the land cover and land use grids. 
Instead of downscaling the high-resolution 10 m classified categories onto a coarser 30 m grid using a majority rule, this procedure calculates a distinct, model-based baseline LST estimate directly for each individual 10 m pixel. The categorical information is therefore retained for interpretation, while the thermal surface remains defined by the hierarchical model.

\begin{table}
    \caption{Summary statistics of the posterior mean of $\alpha+u(\boldsymbol{s})$ by 2023 land cover categories (°C). Categories are displayed in descending order according to their mean, from the warmest to the coolest}
    \centering
    \normalsize
    \begin{tabular}{lccccc}
         \toprule
\textbf{Land Cover} & \textbf{Mean} & \textbf{SD} & \textbf{Median} & \textbf{2.5\%} & \textbf{97.5\%} \\
\midrule
Impermeable surfaces             & 42.09 & 2.55 & 42.17 & 36.01 & 46.84 \\
Permanent herbaceous vegetation  & 39.77 & 2.90 & 40.13 & 33.65 & 44.91 \\
Periodic herbaceous vegetation   & 39.22 & 2.32 & 39.83 & 33.47 & 42.55 \\
Coniferous forest                & 36.75 & 3.05 & 36.43 & 31.87 & 42.39 \\
Unconsolidated surfaces          & 36.15 & 2.06 & 36.04 & 32.66 & 40.00 \\
Broad-leaved forest              & 35.88 & 3.44 & 35.11 & 30.95 & 42.52 \\
Shrubland                        & 35.82 & 2.22 & 35.65 & 32.09 & 40.68 \\
Permanent water bodies           & 35.23 & 2.22 & 35.41 & 31.05 & 39.33 \\
\bottomrule
    \end{tabular}
    \label{tab:lst_lc}
\end{table} 

\begin{table}
\centering
\caption{Summary statistics of the posterior mean of $\alpha + u(\boldsymbol{s})$ by 2023 land use categories (°C). Categories are displayed in descending order according to their mean, from the warmest to the coolest}
\normalsize
\begin{tabular}{lccccc}
\toprule
\textbf{Land Use} & \textbf{Mean} & \textbf{SD} & \textbf{Median} & \textbf{2.5\%} & \textbf{97.5\%} \\
\midrule
Urban                    & 41.32 & 2.84 & 41.59 & 34.60 & 46.29 \\
Arable crops             & 39.22 & 2.32 & 39.83 & 33.47 & 42.55 \\
Forage crops             & 38.28 & 2.41 & 38.62 & 33.43 & 42.27 \\
Other agricultural uses  & 37.26 & 1.25 & 37.37 & 34.76 & 39.76 \\
Permanent crops          & 36.28 & 1.61 & 36.11 & 33.62 & 39.05 \\
Other non-economic uses  & 36.07 & 2.34 & 35.80 & 32.27 & 40.77 \\
Water uses               & 35.08 & 2.19 & 35.24 & 31.02 & 39.03 \\
Forestry use             & 33.57 & 1.87 & 33.27 & 30.71 & 37.86 \\
\bottomrule
\end{tabular}
\label{tab:lst_lu}
\end{table}

Tables~\ref{tab:lst_lc} and~\ref{tab:lst_lu} report spatial summaries of the estimated baseline field by surface type. Baseline LST differs substantially across classes. Impermeable surfaces have the highest mean baseline temperature, at 42.09~$^\circ$C, followed by permanent and periodic herbaceous vegetation. Broad-leaved forests, shrublands, and permanent water bodies have lower mean values, generally between 35 and 36~$^\circ$C. Land use summaries show a similar pattern: urbanised areas have the highest mean baseline LST, at 41.32~$^\circ$C, whereas forestry and hydrological uses have the lowest values. These differences are consistent with the thermal properties of built surfaces, vegetation, and water bodies, and indicate that the latent baseline field captures systematic environmental variation.

\begin{figure}[htbp!]
  \centering
  \includegraphics[width=\textwidth]{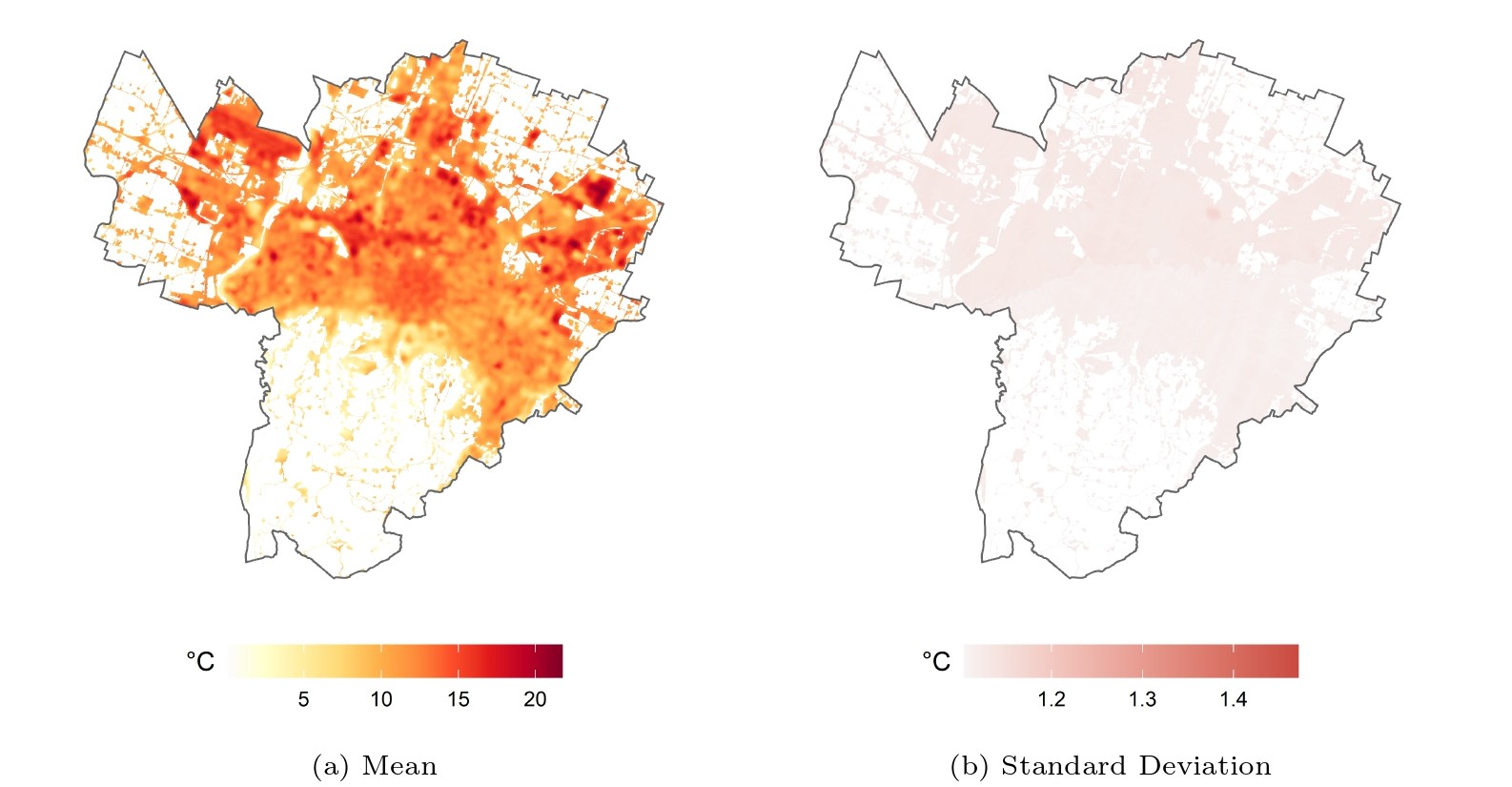}
  \caption{Posterior summaries of surface urban heat island (°C), including a) mean and b) standard deviation}
  \label{fig:SUHI}
\end{figure}

\subsection{Surface urban heat island across land cover classes}

Figure~\ref{fig:SUHI} presents the posterior mean and standard deviation of the SUHI (Equations \ref{eq:SUHI-est}-\ref{eq:SUHI-sd}). The mean map shows clear spatial heterogeneity across the municipality, with values ranging from 0.1 to 21.8~$^\circ$C. The highest SUHI levels occur mainly within the urban core, while peripheral and densely vegetated areas show lower values. Posterior standard deviations remain low across most of the domain, indicating that the main spatial contrasts are estimated with relatively high precision.

Table~\ref{tab:suhi_land_cover} summarises SUHI by urban land cover class. Impermeable surfaces have the highest mean SUHI, at 12.01~$^\circ$C, reflecting the thermal excess associated with built-up areas. Permanent herbaceous vegetation also shows a relatively high mean SUHI, at 10.93~$^\circ$C. Forested classes have lower mean values, ranging from 8.16 to 8.52~$^\circ$C, while permanent water bodies have the lowest mean SUHI levels, at 6.00~$^\circ$C. These patterns indicate that SUHI is strongly related to surface properties, with lower values associated with covers that support shading, higher reflectance, or evaporative cooling. 

The definition of the non-urban reference temperature is a key methodological choice in SUHI estimation, as it directly determines the magnitude of the estimated thermal contrast \parencite{MartinVide2015}. To assess the sensitivity of the results to this choice, alternative non-urban reference temperatures were computed using the mean, 5th and 10th percentiles of the posterior distribution of $B(\boldsymbol{s})$ within each elevation band, alongside the minimum-based reference adopted in the main analysis (Table OR4.1 in Online Resource 4). The overall spatial pattern of SUHI remained consistent across the different reference definitions, although the estimated intensities varied, with lower values obtained using percentile- or mean-based references. These findings suggest that the spatial distribution of SUHI is relatively robust to the choice of reference temperature, whereas its absolute magnitude remains sensitive to the selected non-urban baseline.
\begin{table}
    \centering
    \caption{Summary statistics of the surface urban heat island by 2023 urban land cover category (°C). Categories are displayed in descending order according to their mean, from the warmest to the coolest}
    \normalsize
    \begin{tabular}{lccccc}
        \toprule
        \textbf{Land Cover} & \textbf{Mean} & \textbf{SD} & \textbf{Median} & \textbf{2.5\%} & \textbf{97.5\%} \\
        \midrule
        Impermeable surfaces            & 12.01 & 2.51 & 12.10 & 6.01 & 16.67 \\
        Permanent herbaceous vegetation & 10.93 & 2.37 & 11.14 & 5.28 & 15.19 \\
        Shrubland                       & 8.65 & 1.94 & 8.88 & 4.07 & 12.47 \\
        Broad-leaved forest             & 8.52 & 2.79 & 8.95 & 2.78 & 13.12 \\
        Coniferous forest               & 8.16 & 2.63 & 8.33 & 3.15 & 12.79 \\
        Permanent water bodies          & 6.00 & 1.90 & 6.01 & 2.88 & 10.28 \\
        \bottomrule
    \end{tabular}
    \label{tab:suhi_land_cover}
\end{table}

\subsection{Spatial variation in LST trends}

Long-term changes in summer LST were assessed using the total linear trend component, $T(\boldsymbol{s})=\beta+v(\boldsymbol{s})$, which represents the local annual rate of temperature change. The posterior mean surface shows a positive trend across most of the study region, with only a few small areas displaying negative rates (Figure~\ref{fig:posterior_mean_u_v}). This spatial pattern indicates widespread summer warming over the 13-year period, with local differences in trend magnitude.

Tables~\ref{tab:lst_trend_land_cover} and~\ref{tab:lst_trend_land_use} report trend summaries by land cover and land use class. Among land cover types, periodic herbaceous vegetation has the highest mean rate of change, at 0.22~$^\circ$C per year, exceeding the rates estimated for permanent water bodies, coniferous forests, and broad-leaved forests. Impermeable surfaces and shrublands show lower mean trends of approximately 0.11~$^\circ$C per year. This should be interpreted as a stronger estimated rate of change from a cooler baseline, rather than as higher absolute surface temperature over water, matching documented global trends of rapid aquatic surface warming \parencite{OReilly2015-yv}. The pattern may reflect the sensitivity of small water bodies to summer meteorological variability, water depth, and shoreline microclimates \parencite{Hathway2012-jk}. Furthermore, because this class covers a limited area, its trend estimate is inherently more susceptible to localized land-cover anomalies than the macro-estimates derived for widespread impervious surfaces \parencite{ZHOU201154}. For land use classes, arable crop areas show the fastest warming, while urban areas have a mean trend of 0.12~$^\circ$C per year. These differences indicate that temporal change varies across landscape types and does not increase in direct proportion to baseline urban heat intensity.

\begin{table}
    \centering
    \caption{Summary statistics of the posterior mean of $\beta + v(\boldsymbol{s})$ by 2023 land cover category (°C/year). Categories are displayed in descending order according to their mean, from the warmest to the coolest}
    \normalsize
    \begin{tabular}{lccccc}
        \toprule
        \textbf{Land Cover} & \textbf{Mean} & \textbf{SD} & \textbf{Median} & \textbf{2.5\%} & \textbf{97.5\%} \\
        \midrule
        Periodic herbaceous vegetation  & 0.22 & 0.15 & 0.21 & -0.06 & 0.51 \\
        Permanent water bodies          & 0.17 & 0.10 & 0.16 & -0.01 & 0.41 \\
        Coniferous forest               & 0.15 & 0.08 & 0.15 &  0.00 & 0.28 \\
        Permanent herbaceous vegetation & 0.14 & 0.11 & 0.14 & -0.07 & 0.38 \\
        Broad-leaved forest             & 0.14 & 0.08 & 0.14 & -0.03 & 0.29 \\
        Impermeable surfaces            & 0.11 & 0.10 & 0.11 & -0.08 & 0.33 \\
        Shrubland                       & 0.11 & 0.10 & 0.11 & -0.09 & 0.31 \\
        Unconsolidated surfaces         & 0.08 & 0.07 & 0.08 & -0.06 & 0.21 \\
        \bottomrule
    \end{tabular}
    \label{tab:lst_trend_land_cover}
\end{table}

\begin{table}
    \centering
    \caption{Summary statistics of the posterior mean of $\beta + v(\boldsymbol{s})$ by 2023 land use category (°C/year). Categories are displayed in descending order according to their mean, from the warmest to the coolest}
    \normalsize
    \begin{tabular}{lccccc}
        \toprule
        \textbf{Land Use} & \textbf{Mean} & \textbf{SD} & \textbf{Median} & \textbf{2.5\%} & \textbf{97.5\%} \\
        \midrule
        Arable crops             & 0.22 & 0.15 & 0.21 & -0.06 & 0.51 \\
        Water uses               & 0.17 & 0.10 & 0.16 & -0.01 & 0.41 \\
        Forage crops             & 0.16 & 0.13 & 0.16 & -0.09 & 0.42 \\
        Forestry use             & 0.14 & 0.07 & 0.15 & -0.01 & 0.27 \\
        Urban                    & 0.12 & 0.10 & 0.12 & -0.07 & 0.33 \\
        Other agricultural uses  & 0.12 & 0.08 & 0.12 & -0.02 & 0.28 \\
        Other non-economic uses  & 0.11 & 0.10 & 0.11 & -0.07 & 0.33 \\
        Permanent crops          & 0.08 & 0.07 & 0.07 & -0.03 & 0.27 \\
        \bottomrule
    \end{tabular}
    \label{tab:lst_trend_land_use}
\end{table}

\subsection{Excursion sets and local landscape changes}

\begin{figure}[htbp!]
  \centering
  \includegraphics[width=\textwidth]{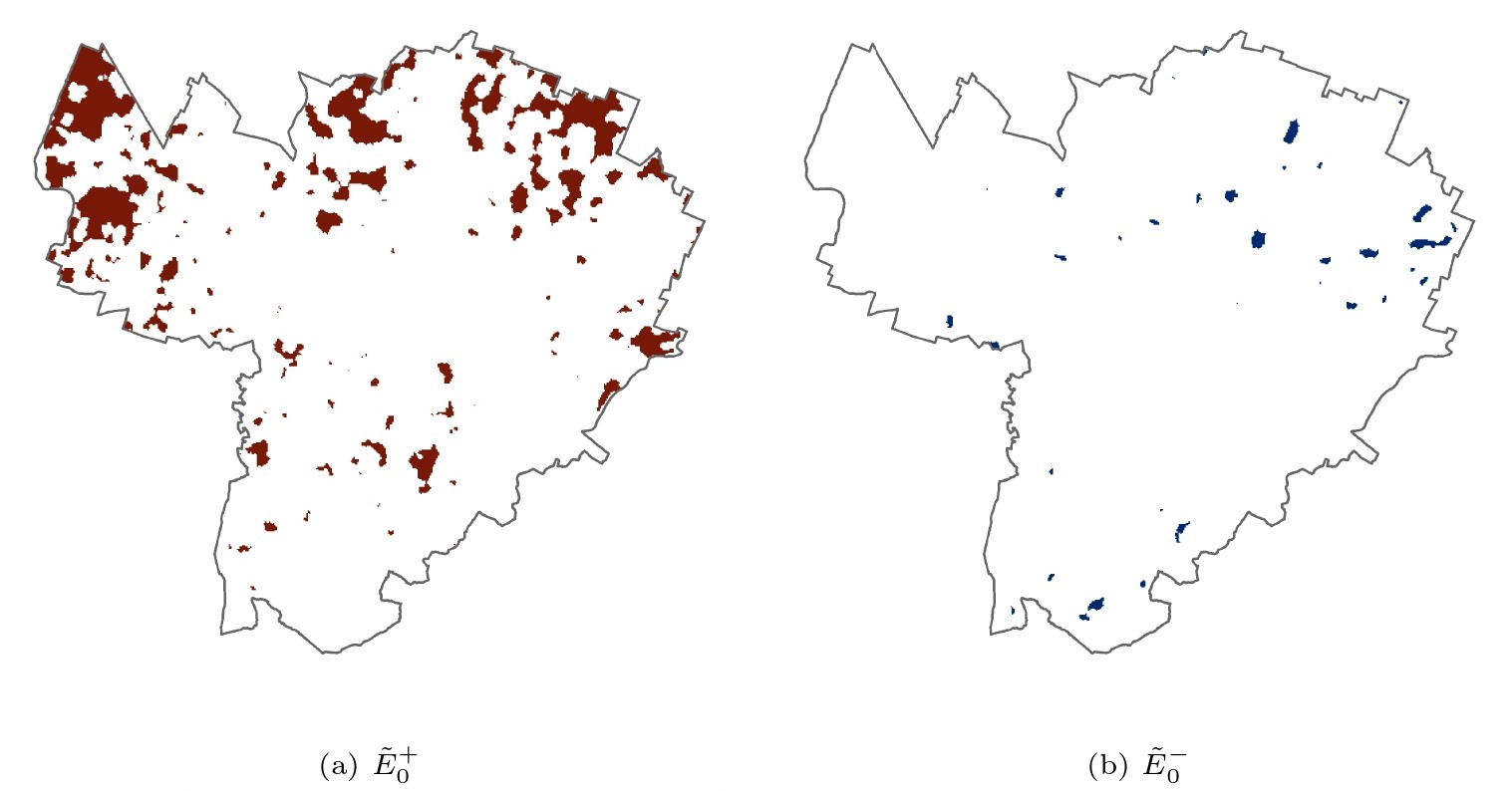}
  \caption{95\% excursion sets identifying a) regions of above-average warming, defined by $v(\boldsymbol{s})>0$, and b) regions of absolute cooling, defined by $T(\boldsymbol{s})<0$}
  \label{fig:exc_ALL}
\end{figure}

Posterior excursion sets were computed to identify regions where temporal trends are statistically supported under simultaneous probability control. Figure~\ref{fig:exc_ALL} shows the 95\% excursion boundaries. The positive excursion set $\tilde{E}^{+}_{0}$ (Equation \ref{eq:pos-exc}) identifies areas where the local trend component $v(\boldsymbol{s})$ is greater than zero, indicating warming rates above the municipal average $\beta$. These areas are spatially extensive and more connected in the northern part of the municipality. In contrast, the negative excursion set $\tilde{E}^{-}_{0}$ (Equation \ref{eq:neg-exc}) identifies only small and fragmented clusters where the total trend $T(\boldsymbol{s})$ is below zero. This asymmetry reflects the positive global trend and indicates that statistically supported cooling is spatially limited.

To aid interpretation of these temporal patterns, selected areas within the excursion sets were examined using high-resolution imagery from Google Earth Pro \parencite{Google2024GoogleSoftware}. Historical images from June 2014 and June 2025 were compared to identify physical landscape changes; 2014 was used as the earliest reliable reference year because imagery for 2013 was unavailable. The case studies are reported in Online Resource 3. Within the warming excursion sets, increases in LST are often associated with new built structures, increased soil sealing, canopy removal, or degradation of existing urban green space. In the smaller cooling excursion sets, local decreases in LST correspond to vegetation maturation or densification, or to land-use interventions that increased shading or surface reflectance.

\subsection{Prior sensitivity analysis}

A prior sensitivity analysis was conducted to assess the influence of PC prior hyperparameters for the spatial range components (Equations \ref{eq:pc_u} and \ref{eq:pc_v}). The use of PC priors facilitates this assessment because their hyperparameters can be elicited through interpretable probability statements, allowing alternative degrees of prior regularisation to be specified and compared systematically. Details are provided in Online Resource 4. 

The global fixed effects are stable across all tested prior configurations. The posterior mean of the global trend $\beta$ remains positive and stable, varying within a narrow range from 0.13 to 0.15~$^\circ$C per year. This stability indicates that domain-wide estimates are mainly data-driven and are not strongly affected by the spatial covariance prior specification.

The spatial hyperparameters are more sensitive to prior bounds, as expected. Models with moderate and structurally informed prior ranges produce posterior ranges and marginal standard deviations consistent with the baseline specification. In contrast, extreme prior combinations yield less plausible parameter estimates. For example, overly large prior ranges for both spatial components produce an artificial contraction of the estimated practical range for $u(\boldsymbol{s})$ together with inflated marginal standard deviations. This behaviour suggests weak identifiability and local overfitting when the priors are too diffuse. Information criteria provide further support for the baseline model, DIC and WAIC are lowest and stable under intermediate prior specifications, but increase under extreme or misspecified bounds. The sensitivity analysis therefore supports the selected baseline prior specification as a reasonable balance between parameter identifiability, stable global estimates, and predictive performance.

\section{Discussion}\label{sec:discussion}

\subsection{Main findings and interpretation}

The results show that baseline surface temperature, SUHI, and temporal warming are spatially structured but not identical phenomena. Impervious and urban surfaces exhibited the highest baseline LST and the strongest SUHI, whereas forested areas and water bodies were associated with lower surface temperatures. In contrast, the largest estimated temporal trends were not restricted to the areas with the highest baseline LST. This distinction indicates that present-day surface thermal intensity and recent summer LST change are governed by partly different spatial processes. Posterior excursion sets further support this interpretation by identifying regions where warming above the municipal average or absolute cooling is supported under simultaneous probability control.

\subsection{Modelling choices and interpretation of posterior quantities} 
To retain the distinction between the spatial and temporal components, the model was kept parsimonious, while remaining computationally feasible for a large remote-sensing dataset. The temporal effect was specified as linear, and the residual precision was assumed to be constant across space and time. This formulation was chosen since the main objective of the study was to identify broad spatial patterns and long-term changes in surface temperature, and not to reproduce the short-term fluctuations observed in individual satellite images due to transient meteorological conditions or acquisition-specific variability. More complex temporal or residual structures could improve predictive performance in some applications, but they would reduce the direct interpretability of the model components and increase computational cost. In the final model configuration, estimation required approximately 17 minutes of wall-clock time and 38 GB of peak memory usage, indicating that computational cost remains substantial even under the parsimonious specification. 

Another methodological choice concerned the treatment of land cover and land use, which were not included as covariates in the linear predictor. This was a deliberate modelling choice since, if included directly, these variables would make the spatial fields $u(\boldsymbol{s})$ and $v(\boldsymbol{s})$ represent residual spatial variation after adjustment for selected surface characteristics. The objective here was instead to estimate the overall spatial structure of the LST field and then examine its empirical association with land cover and land use in a second stage. The resulting class summaries should therefore be interpreted as exploratory associations, not as causal or regression-adjusted effects of land cover or land use on LST.

A further modelling decision involved the definition of the non-urban reference temperature in SUHI estimation. In this study, SUHI was calculated as the difference between the posterior baseline LST and the minimum value among non-urban pixels within the same elevation band (Equation~\ref{eq:SUHI}), in order to measure surface thermal excess relative to the coolest modelled non-urban conditions at comparable elevations, rather than to average background temperatures. The sensitivity analysis indicates that, although the reference definition influences the absolute magnitude of SUHI, its overall spatial distribution remains relatively stable across alternative definitions (details in Online Resource 4).

\subsection{Model sensitivity and limitations}

PC priors were used to regularise the SPDE fields and to express assumptions about spatial range and marginal variability through interpretable probability statements. The sensitivity analysis indicates that the main empirical findings remain consistent under moderate prior choices, supporting the robustness of the baseline prior specification.

Despite this robustness, some limitations should be considered when interpreting the results. First, land cover and land use were treated as static, using the 2023 classifications throughout the study period. This may introduce temporal mismatch when interpreting earlier observations, although broad land cover and land use transitions appear limited over the period analysed (see Online Resource 1, Tables OR1.2 and OR1.3). Second, the analysis involves a change-of-support issue, since LST observations and land-cover and land-use data are defined on different spatial supports. Although the Landsat 8 surface temperature product is distributed at 30 m resolution, the underlying thermal measurements have a native spatial resolution of approximately 100 m. Projecting posterior summaries onto the land-cover grid provides model-based estimates of the latent spatial field at those locations, but neither resampling nor posterior projection creates additional independent thermal information at finer spatial scales.

Additional limitations arise from the characteristics of the satellite observations.  Landsat 8 revisits the same location every 16 days, providing observations at a fixed daytime overpass time. This temporal sampling represents a trade-off associated with the choice of the dataset: products with higher temporal frequency, such as MODIS, provide near-daily observations but at substantially coarser spatial resolution, whereas Landsat allows urban thermal patterns to be resolved at much finer spatial detail. The analysis is therefore focused on daytime surface heat patterns and does not address nocturnal UHI dynamics, which may differ from those observed during the day. 

Finally, short-term weather conditions may also affect the LST measured in individual Landsat images. Although the model is designed to capture persistent spatial patterns and long-term temporal changes, acquisition-specific meteorological conditions may still influence the observations. In particular, local effects related to recent rainfall, soil moisture, or vegetation conditions may introduce temperature variations that are not explicitly represented by the model.

\section{Conclusions}
\label{sec:conclusions}

This study developed a Bayesian spatio-temporal framework for analysing SUHI patterns and their evolution over time using satellite-derived LST. The model combines a spatially varying coefficient specification with the SPDE--INLA framework, allowing persistent spatial differences in baseline LST and spatial heterogeneity in temporal trends to be estimated within a single probabilistic model. From a methodological perspective, the proposed framework provides a computationally feasible approach to modelling SUHI as a stochastic spatio-temporal process and analysing large remote-sensing datasets while preserving the interpretability of the latent components and quantifying associated uncertainty. In particular, the use of posterior excursion sets extends the analysis beyond pointwise estimates by identifying spatial regions where warming or cooling exceeds specified thresholds with simultaneous probability control.

Future work could extend the framework in several directions. Remote-sensing products such as MODIS Terra/Aqua and ASTER could be used to explore the trade-off between temporal frequency and spatial resolution \parencite{Wan2019MODISGuide, Yamaguchi1998OverviewASTER}. More frequent observations could also support richer temporal specifications, including non-linear trends, seasonal variability, and responses to extreme events such as heatwaves. Further extensions could examine the relationships between the estimated thermal patterns and urban morphology, vegetation characteristics, and socio-economic conditions, providing a basis for evaluating potential heat-mitigation strategies such as urban greening and reforestation. Finally, integrating satellite-derived LST with in situ meteorological observations or near-surface air temperature estimates could broaden the analysis toward local atmospheric conditions and human heat exposure, complementing the surface-based perspective adopted in this work.

\backmatter

\bmhead{Supplementary information} Ancillary material is provided and includes the following additional online resources:
\begin{itemize}
    \item Online Resource 1: Exploratory analyses;
    \item Online Resource 2: Additional results;
    \item Online Resource 3: Warming and cooling posterior excursion sets;
    \item Online Resource 4: Sensitivity analyses.
\end{itemize}


\printbibliography% common bib file

\section*{Statements and Declarations}

\bmhead{Funding}
This work was supported by the project \textit{Cobenefici di salute ed equità a supporto dei piani di risposta ai cambiamenti climatici in Italia} [Health and Equity Co-Benefits in Support of Plans to Respond to Climate Change in Italy], funded by the Italian National Plan for Complementary Investments of the Ministry of Health, project code PREV-A-2022-12376994 (CUP J55I22004450001).

\bmhead{Competing Interests}
The authors have no relevant financial or non-financial interests to disclose.

\bmhead{Author Contributions}
All authors contributed to the study conception and design. Material preparation, data collection and analysis were performed by Marika D'Agostini. The statistical model was developed by Marika D'Agostini with the supervision of Massimo Ventrucci. The first draft of the manuscript was written by Marika D'Agostini, and all authors commented on previous versions of the manuscript. All authors read and approved the final manuscript.

\bmhead{Data availability}
A subset of the land surface temperature data used in this study is available at \url{https://github.com/marika-dagostini/bhm-svc-suhi}. Owing to storage constraints associated with the full remote-sensing dataset, only data for 2013-2016 are provided. This subset allows the accompanying code to be tested and the main analysis workflow to be reproduced. The complete Landsat 8 dataset can be obtained from the Google Earth Engine Data Catalog. Land-use and land-cover data are publicly available through the National Environmental Information System of the Italian Institute for Environmental Protection and Research (ISPRA), while elevation data are available from the Geoportal of the Emilia-Romagna Region. The corresponding source URLs are cited in the main text.

\bmhead{Code availability} 
All R scripts required to reproduce the computational workflow are available at \url{https://github.com/marika-dagostini/bhm-svc-suhi}. The repository also includes a web map for the interactive visualisation of the model outputs, available at \url{https://marika-dagostini.github.io/bhm-svc-suhi/}.

\bmhead{Ethics approval} Not applicable

\bmhead{Consent to participate} Not applicable

\end{document}